# Uranium distribution in the Variscan Basement of Northeastern Sardinia

Kaçeli Xhixha M.[a,b], Albèri M. [c,e], Baldoncini M. [b,c,e], Bezzon G.P. [d], Buso G.P.[d], Callegari I. [b,d], Casini L.[f], Cuccuru S.[f], Fiorentini G.[c,e], Guastaldi E.[b,g], Mantovani F.[c,e], Mou L.[d], Oggiano G. [f], Puccini A.[f], Rossi Alvarez C. [d], Strati V.[b,c,d*], Xhixha G. [b,d], Zanon A[d].

[a] University "Aleksandër Moisiu" Durrës, Department of Engineering Sciences, Faculty of Professional Studies, Str. Currila 1, 2000 - Durrës, Albania.
[b] GeoExplorer Impresa Sociale S.r.l., Via E. Vezzosi, 15, 52100 - Arezzo, Italy.
[c] University of Ferrara, Department of Physics and Earth Sciences, Via Saragat 1, 44121 - Ferrara, Italy.
[d] INFN, Legnaro National Laboratories, Viale dell'Università, 2 - 35020 Legnaro, Padua, Italy.
[e] INFN, Ferrara Section, Via Saragat 1, 44121 - Ferrara, Italy.
[f] University of Sassari, Nature and Environment Department, via Piandanna 4 - 07100, Sassari, Italy.
[g] University of Siena, Center for GeoTechonologies, Via Vetri Vecchi 34 - 52027 San Giovanni Valdarno, Arezzo, Italy.

* Corresponding author: Virginia Strati, Department of Physics and Earth Sciences, University of Ferrara, Via Saragat, 1 - 44122 Ferrara, Italy. Phone: +39 3489356603. Email: strati@fe.infn.it

# Abstract

We present a detailed map of the uranium distribution and its uncertainties in the Variscan Basement of Northeastern Sardinia (VBNS) at a scale 1:100,000. An area of 2100 km$^2$ was investigated by means of 535 data points obtained from laboratory and *in situ* gamma-ray spectrometry measurements. These data volume corresponds to the highest sampling density of the European Variscides, aimed at studying the genetic processes of the upper crust potentially triggered by an enrichment of radiogenic heat-producing elements. For the first time the Kriging with Variance of Measurement Error method was used to assign weights to the input data which are based on the degree of confidence associated to the measurements obtained with different gamma-ray spectrometry techniques. A detailed tuning of the model parameters for the adopted Experimental Semi-Variogram led to identify a maximum distance of spatial variability coherent to the observed tendency of the experimental data. We demonstrate that the obtained uranium distribution in the VBNS, characterized by several calc-alkaline plutons emplaced within migmatitic massifs and amphibolite-facies metamorphic rocks, is an excellent benchmark for the study of 'hot' collisional chains. The uranium map of VBNS, and in particular the Arzachena minor pluton, confirms the emplacement model based on the recognition of the different petrological associations characterizing the Variscan magmatic processes in the Late Paleozoic. Furthermore, the presented model of the uranium content of the geological bedrock is a potential baseline for future mapping of radon-prone areas.

## 1. Introduction

The Variscan Basement of Northeastern Sardinia (VBNS) is a benchmark for the study of 'hot' collisional chains characterized by a high temperature-low pressure (HT-LP) gradient. Several processes might have enhanced the Variscan geotherm, such as i) shear heating **(Casini, Cuccuru, Puccini, Oggiano, & Rossi, 2015; Maino et al., 2015)**, ii) the advection of hot, mantle-derived melts, iii) the break-off of the mantle lithosphere **(Li, Faure, & Lin, 2014)** or iv) the selective enrichment of radiogenic heat-producing elements, such as U, Th and K, in the crust **(Lexa et al., 2011)**. The widespread late-Variscan magmatism in the Corsica-Sardinia Batholith (C-SB) has been occasionally explained in terms of enhanced radiogenic heating. The models that describe the efficiency of the processes depend strongly on the uranium content of the fertile crust **(Bea, 2012)**. The heterogeneous distribution of uranium throughout the VBNS might be a proxy for investigating the applicability of thermal models based on the selective enrichment of radiogenic elements in the crust **(Mohamud, Cózar, Rodrigo-Naharro, & Pérez del Villar, 2015; Tartèse, Boulvais, Poujol, & Vigneresse, 2011)**. In addition, the outcrops in VBNS are the most accessible intrusive bodies for studying the geoneutrino signal in the Borexino experiment **(Borexino Collaboration, 2015)**, which is particularly sensitive to the U and Th contents and distributions in the Variscan continental crust **(Coltorti et al., 2011)**.

Finally, this study has implications related to the public health, as recent investigations **(Bochicchio et al., 2005)** showed that the Sardinia region is characterized by high values of radon gas, monitored in 124 dwellings. Since for good bedrock exposure, as in the case of VBNS, the correlation between indoor radon concentrations and uranium content of the underlying rocks increases, the results of this study potentially constitute a baseline for future mapping of radon-prone areas.

In this paper, we present a map of the eU distribution (this notation indicates equivalent uranium, as we assume secular equilibrium in the $^{238}$U decay chain) in the VBNS at a scale of 1:100,000 as support for further studies regarding the main geophysical, geochemical and geodynamic features of the continental crust in this region. This study is included in the framework of a research project which has already led to the realization of the total natural radioactivity map of the Tuscany region **(Callegari et al., 2013)** and Veneto region **(Strati et al., 2015)**.

The spatial model, together with its uncertainties, was obtained using the Kriging with Variance of Measurement Error method for 535 gamma-ray spectrometry measurements. The eU distribution is discussed in the geodynamic framework proposed by **(Casini, Cuccuru, Puccini, et al., 2015)**, taking into account the petrological features of the C-SB, the compositional variation and the emplacement timing of Variscan granitoids. Finally, we focus on the eU distribution measured in the granitoids of the Arzachena pluton, linking it to the emplacement mechanism proposed in **(Casini, Cuccuru, Maino, Oggiano, & Tiepolo, 2012)**.

## 2. Geological setting

The Variscan belt of Western Europe resulted from the collision of Northern Gondwana and Laurussia in a time interval spanning from the Late Devonian to the Early Permian (~380-270 Ma). The European Variscan crust, including the Corsica-Sardinia massif, experienced several episodes of plutonic and volcanic activity with different petrochemical affinities **(Finger, Roberts, Haunschmid, Schermaier, & Steyrer, 1997) (Corsini & Rolland, 2009).**

The C-SB, with its ca. 12,000 km$^2$ area, constitutes one of the largest batholiths in south-western Europe, emplaced in approximately 40 Ma (Late Mississippian Pennsylvanian-Early Permian). Three main magmatic suites can be recognized: a magnesium-potassium complex exposed only in northern Corsica, a peraluminous calc-alkaline complex **(Casini et **al., 2012; Rossi & Cocherie, 1991)** and finally, a late to post-orogenic alkaline suite in the VBNS **(Bonin, 2007)**. The 2,100 km$^2$ of the VBNS (**Figure 1**) are characterized by several calc-alkaline plutons and a few minor alkaline complexes emplaced within migmatites and amphibolite-facies of metamorphic rocks **(Casini, Cuccuru, Maino, et al., 2015)**. According to **(Cruciani, Franceschelli, Massonne, Carosi, & Montomoli, 2013)**, the migmatites outcropping in the study area are related to high-pressure metamorphism occurred in the Internal Nappes of the Sardinia Variscides at the age of the initial continent–continent collision.

Both migmatites and the calc-alkaline plutons have been interpreted in terms of extensive crustal melting related to the establishment of an anomalous thermal gradient **(Anderson, 2006)**. One of the main contributions to the high geothermal gradient, which is required to induce anatexis processes, originates from the enrichment of radiogenic heat-producing elements caused by several genetic processes, such as subduction of continental crust, crust-scale migmatization **(Li et al., 2014),** melt dehydration and segregation **(Gerdes, Worner, & Henk, 2000)**. Understanding heat production and transfer mechanisms is relevant for modeling thermal-kinematic and exhumation processes **(Lexa et al., 2011)**.

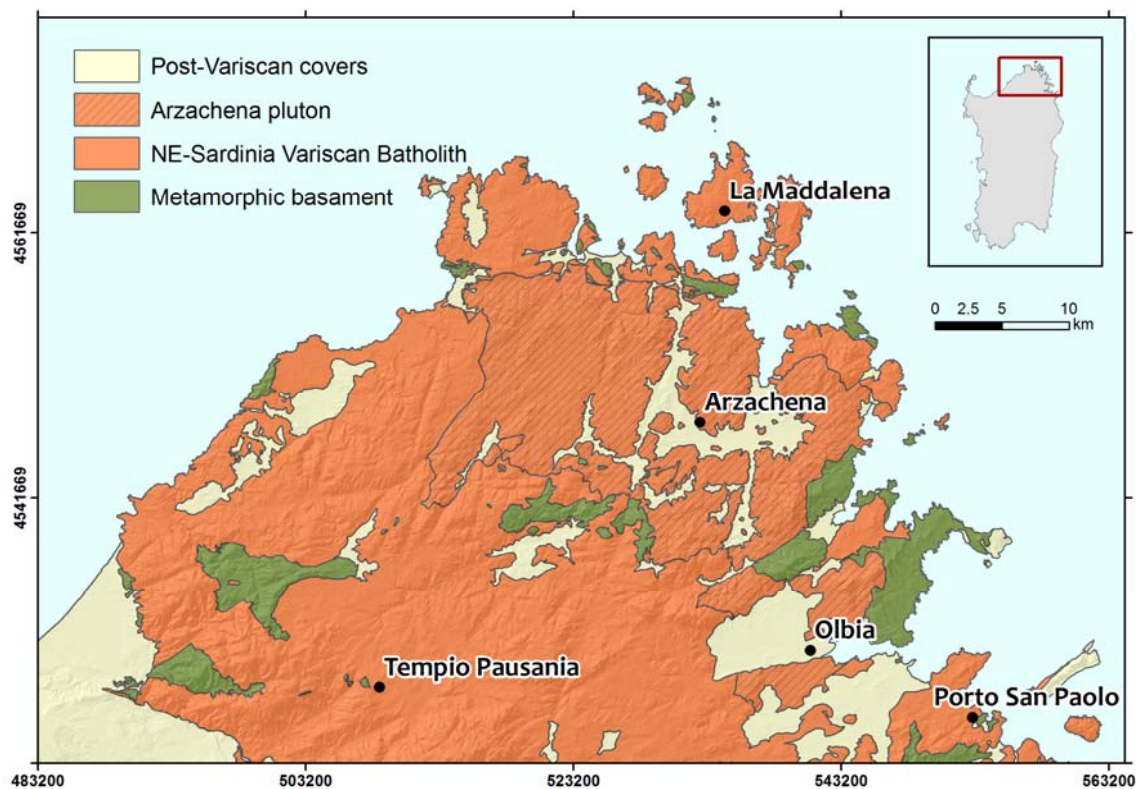

**Figure 1** - Geological sketch map of the VBSN (cartographic reference system WGS84 UTM ZONE 32N), modified from (Casini, Cuccuru, Maino, et al., 2015).

## 3. Methods

The 535 input data points used for producing the Main Map come from 167 rock sample measurements and 368 *in situ* measurements. The surveys were planned based on the Geological map of Sardinia at a scale of 1:200,000 **(Barca et al., 1996)** and the structural map of Variscan Northern Sardinia at a scale of 1:100,000 **(Casini, Cuccuru, Maino, et al., 2015).**

The eU abundances in the rock samples collected from fresh outcrops were measured using the MCA-Rad system, a gamma-ray spectrometer equipped with two HPGe p-type detectors having a 60% relative coaxial efficiency and an energy resolution of approximately 1.9 keV at 1332.5 keV ($^{60}$Co) **(G. Xhixha et al., 2013)**. The MCA_Rad system is accurately shielded with 10-cm thickness of copper and 10-cm thickness of lead by reducing the laboratory background of approximately two orders of magnitude. The absolute peak energy efficiency of the MCA_Rad system is calibrated using certified reference materials (RGK_1, RGU_1 and RGTh_1) traceable by the International Atomic Energy Agency (IAEA) (Gerti Xhixha et al., 2015). The total uncertainty for the absolute peak energy efficiency is estimated to be less than 5%. Prior to measurement, each rock sample was crushed, homogenized and sealed in a cylindrical polycarbonate container of 180 cm$^3$ volume. Then, the samples were left undisturbed for at least four weeks to establish radioactive equilibrium in the $^{226}$Ra decay chain segment. Each sample was measured for 3600 seconds with a statistical uncertainty generally less than 10 % for eU. Less than 2% of the measurements are below the minimum detection activity of ~2.5 Bq/kg.

In this survey, a total of 368 *in situ* gamma-ray measurements were performed on granitic outcrops using a portable NaI(Tl) detector mounted in a backpack to allow flexible operation **(Figure 2)**, with an acquisition live time of 300 seconds. The NaI(Tl) detector is of a cubic shape (10.2 cm × 10.2 cm × 10.2 cm) and has an energy resolution of 7.3 % at 662 keV($^{137}$Cs) and 5.2% at 1,172 and 1,332 keV ($^{60}$Co). The instrument was calibrated following the method of Full Spectrum Analysis with the Non-Negative Least Squares (FSA-NNLS) constraint, as described in **(Caciolli et al., 2012)**. According to the FSA-NNLS method, each spectrum was reconstructed from a linear combination of standard spectra for $^{238}$U, $^{232}$Th, $^{40}$K, $^{137}$Cs and for the background. The uncertainty of the method is estimated to be 5 % for $^{40}$K and 7 % for $^{232}$Th, with relatively higher uncertainty for $^{238}$U of approximately 15%. In **(Caciolli et al., 2012),** the coefficient of correlation (0.87 ± 0.12) between the eU values obtained by NaI(Tl) and HPGe is compatible with the unity at the 1 sigma level. Despite this strong agreement between the two acquisition methods, it is well known **(IAEA, 2003)** that *in situ* gamma ray measurements are susceptible to many sources of 'noise': the geometry of the investigated area, the presence of atmospheric radon, the soil moisture content, the weathering and the outcrop exposure can affect the gamma signal, decreasing the precision of the survey.

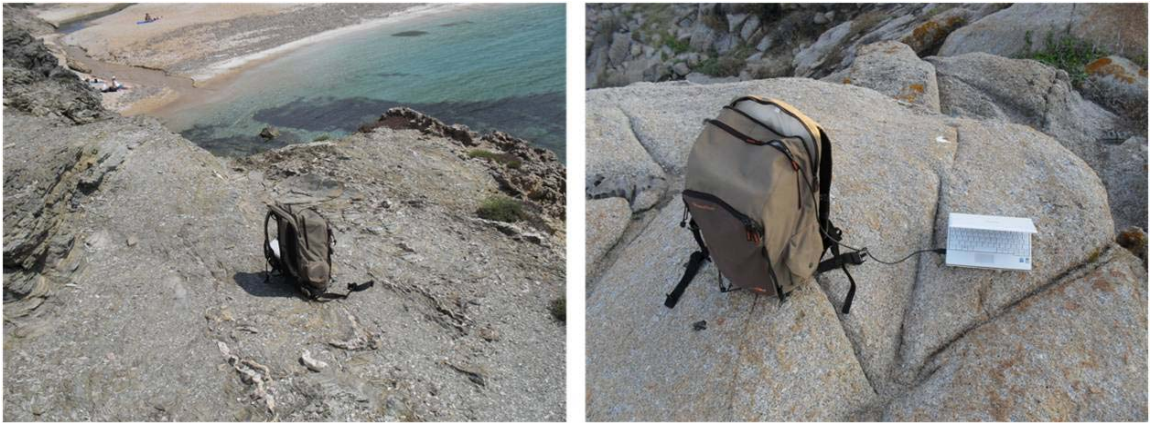

**Figure 2** - Field measurements using the portable NaI (TI) detector.

Because in our study we include the field measurements, relevant precautions were taken to ensure the reliability of our data. Indeed, we avoided acquiring data immediately after rainfall and preferred flat outcrops far from man-made constructions **(Figure 2).**

In the U decay chain, disequilibrium occurs when one or more isotopes are completely or partially removed or added to the system. Because gamma ray spectroscopy is a method that detects the gamma emitter daughters of uranium, secular equilibrium of the decay chains is commonly assumed and is reported as the equivalent uranium (eU).

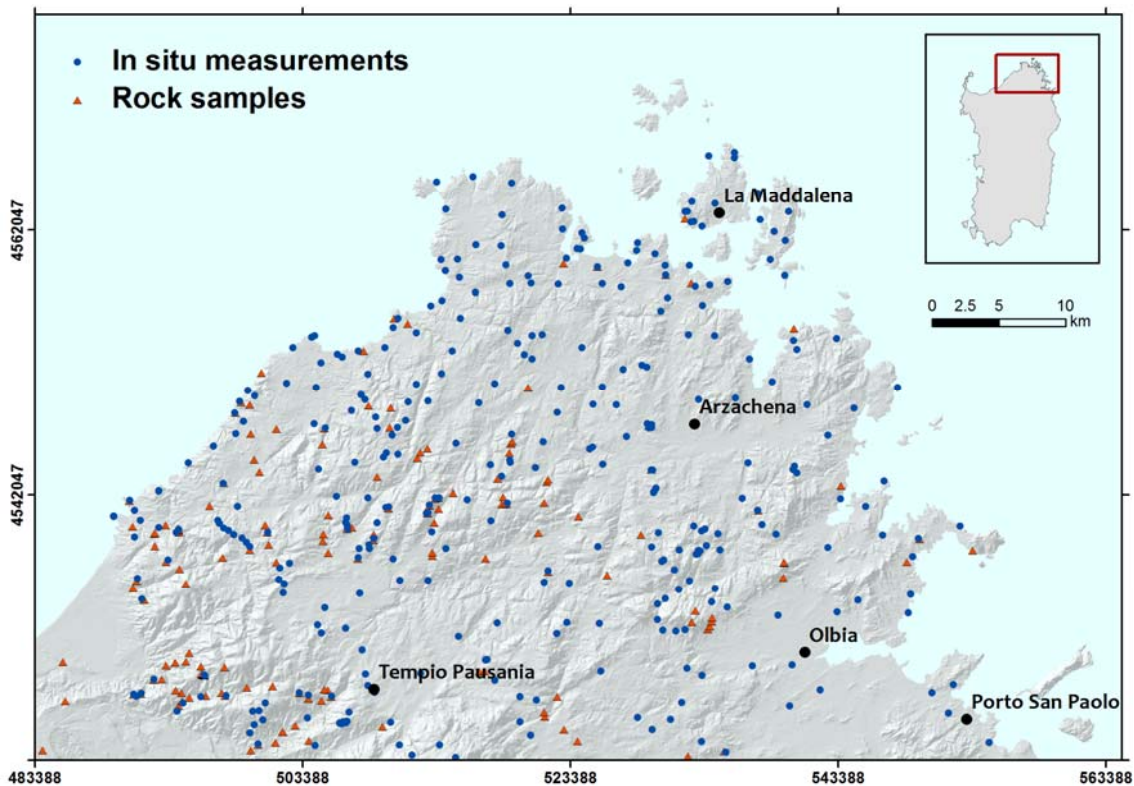

**Figure 3** - The locations of the 167 collected rock samples and the 368 in situ measurements (cartographic reference system WGS84 UTM ZONE 32N).

## 4. Mapping radiometric data

The map of the eU distribution in VBNS, a raster with a 100 m × 100 m spatial resolution, was obtained using all 535 measurements acquired by HPGe and NaI(Tl) using the Kriging method (**Figure 3**). The frequency distribution and the principal statistical parameters of the input data and the output model are reported in the Main Map.

The exploratory statistics analysis highlights that the HPGe and NaI(Tl) data are characterized by frequency distributions **(Figure 4)** described by the statistical parameters reported in **Table 1**. We emphasize that the distributions are affected by a sampling bias because each lithology was not investigated by the same amount of records in the two datasets: this is particularly clear in observing the tails of the distributions. In fact, the quartzites and amphibolites of the metamorphic basement are characterized by the lowest uranium concentrations (< 2 µg/g) and were mainly investigated with HPGe measurements because they are badly exposed and poorly suitable for *in situ* measurements. In contrast, the highest values in **Figure 4** are found in the U-rich monzogranites (~ 9 µg/g) of the La Maddalena pluton **(Casini, Cuccuru, Maino, et al., 2015)**, investigated mostly with in situ surveys.

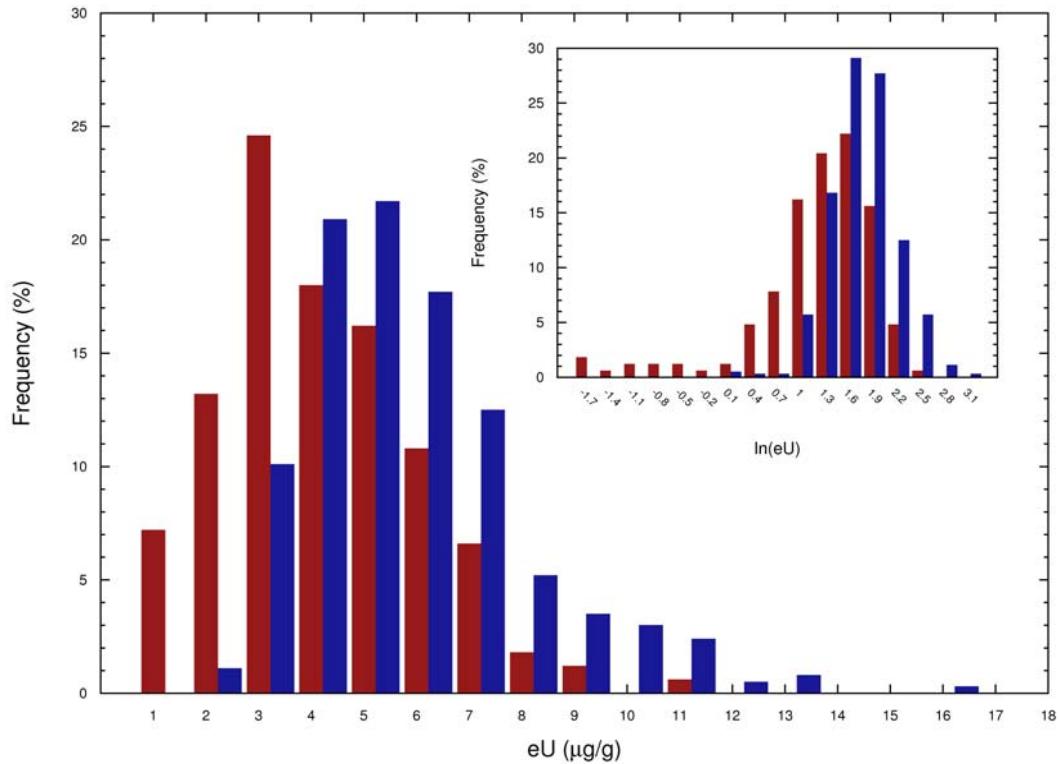

**Figure 4** - The frequency distributions of the eU abundances (µg/g) in rock samples (red) and *in situ* measurements (blue).

Table 1 - Uranium average abundance U (μg/g), standard deviation σ (μg/g) and goodness of fit (normalized χ2) obtained with normal and lognormal probability density functions applied to all data and both datasets (i.e., HPGe and NaI(Tl)).

| Dataset | Number of records | Normal distribution | | Lognormal distribution | |
|---|---|---|---|---|---|
| | | U ± σ (μg/g) | $\chi^2$ | U ± σ (μg/g) | $\chi^2$ |
| HpGe | 167 | 3.5 ± 1.9 | 1.7 | $2.9^{+3.3}_{-1.5}$ | 4.3 |
| NaI(Tl) | 368 | 5.2 ± 2.2 | 6.0 | $4.8^{+2.4}_{-1.6}$ | 0.5 |
| All data | 535 | 4.7 ± 2.3 | 4.5 | $4.1^{+3.3}_{-1.8}$ | 6.1 |

Since the distribution of the input data shows a skewness value close to unity, the study of the spatial variability was performed without any normal transformation of the uranium abundances. A detailed analysis of the directional Experimental Semi-Variograms (ESV) highlighted an isotropic experimental variability without any preferred directions. Therefore, an omnidirectional ESV made up of 9 lags of 2.2 km was computed and modeled using a trial-and-error procedure (**Figure 5**). Since we are interested in small scale variability of U abundances, the parameters used for the ESV modeling were tuned for optimizing the fit in the first lags of the ESV. The nugget effect (1.7 μg$^2$/g$^2$), contributing approximately to 30% of the total amount of spatial variability, and the maximum distance of spatial variability equal to 4.8 km are in excellent agreement with the observed tendency of the experimental data. The goodness of fit of the ESV model was checked via a cross-validation procedure. The results are reported in **Table 2,** together with the parameters of the structures of variability used for the ESV modeling.

Table 2 - Parameters of the structures of variability used for the model fitted on the ESV; results of the cross-validation procedure in terms of the Mean of Standardized Errors (MSE) and the Variance of Standardized Errors (VSE).

| ESV model parameters | | | Cross-validation results | |
|---|---|---|---|---|
| Structures of variability | Range (km) | Sill (μg/g)$^2$ | MSE | VSE |
| Nugget effect model | - | 1.7 | -0.04 | 0.74 |
| First Spherical model | 3.4 | 2.4 | | |
| Second spherical model | 1.4 | 0.8 | | |

The estimation process, performed with Geovariances ISATIS® software, takes into account the overall uncertainties of the two methods of gamma-ray measurements as the known Variance of Measurements Error of the input data. In particular, we considered an overall uncertainty of 5% for each HPGe measurement and a conservative uncertainty of 20% in the case of the NaI(Tl) measurements. This methodology, known as Kriging with Variance of Measurement Error, allows for assigning different weights to the input considering the degree of confidence of the measurements, thus improving the quality of the estimations **(Deraisme & Strydom, 2009)**.

The accuracy of the spatial model in terms of the variance normalized with respect to the estimated values is reported in the Main Map with the contour lines.

The chromatic variations in the color ramp of the legend were assigned to specific values of eU concentration. In particular, they correspond to the $20^{th}$, $35^{th}$, $45^{th}$, $55^{th}$, $65^{th}$, $75^{th}$ and $80^{th}$ percentiles calculated for the entire dataset of 535 measurements.

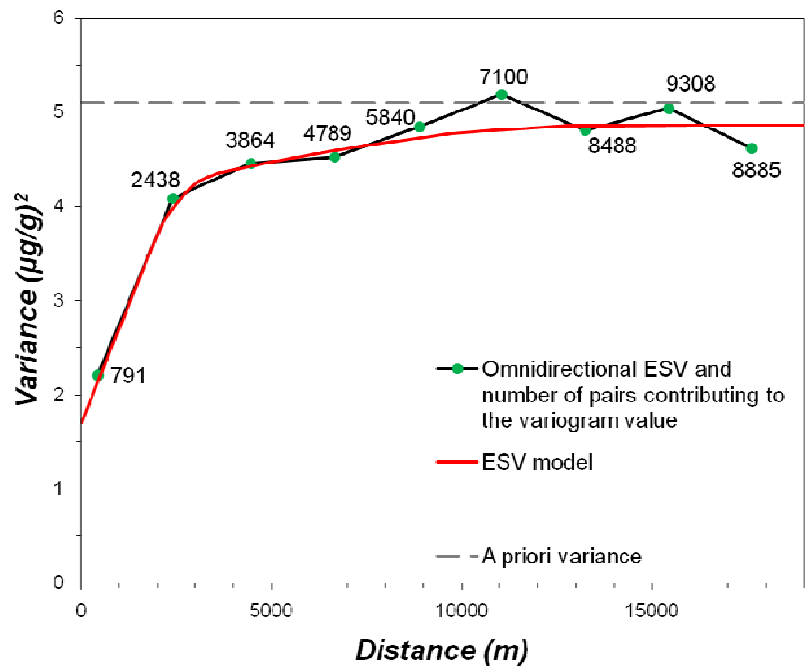

**Figure 5** - Model fitted for the ESV calculated with 9 lags of 2.2 km.

## 5. The Arzachena pluton

In this section, we present a specific focus on the eU distribution in the Arzachena pluton (AZN), a reference calc-alkaline massif of the VBNS.

The AZN is an elongated pluton that consists of three concentric granitic shells **(Oggiano et al., 2005)**. The pluton shows reverse zonation: the more mafic terms (medium-grained granodiorite) are exposed in the core, along the southern margin of the pluton, whereas the external part is composed of felsic rocks, such as porphyritic biotite–monzogranite and fine-grained leuco-monzogranites. This last type of rock represents the more evolved magmatic products and is mainly localized in the peripheral zone of the pluton, exposed to the north **(Casini et al., 2012).**

In **Figure 6**, we use the cartographic boundaries of the three granitic shells to distinguish and analyze the range of eU concentrations of the different petrological associations related to the magmatic differentiation during magma emplacement. The mean and 1σ uncertainty of the input data and of the spatial model are reported for the three rock types. The granodiorites are characterized by lower eU concentrations (3.4 ± 0.4 µg/g). Monzogranites and leuco-monzogranites are enriched in eU content, with average values of 4.3 ± 0.8 µg/g and 4.9 ± 0.7 µg/g, respectively. This distribution defines a clear geochemical trend characterized by eU values that increase in felsic rocks. The eU distribution within the AZN supports a strong positive correlation between the presence of U-bearing accessory minerals (e.g., zircons, monazite and xenotime) and the evolution of magmatic systems **(Pérez-Soba, Villaseca, Orejana, & Jeffries, 2014).** Indeed, melting processes and melt extraction remove a large amount of radioactive elements from the magmatic system that are hosted by accessory minerals in which uranium can substitute other cations with a similar ionic radius and occasionally equal valences (e.g., Ca, trivalent REE, Zr and Y) **(Mohamud et al., 2015; Pagel, 1982; Peiffert, nguyen-Trung, & Cuney, 1996).**

These results confirm the conceptual model proposed in **(Casini et al., 2012),** which explains the distribution of Large-Ion Lithophile Elements (LILE) within AZN in terms of *in situ* progressive differentiation of anatectic melts sourced from a compositionally heterogeneous lower crust. The obtained eU distribution in the AZN agree with the emplacement history and the pluton evolution described in the proposed termo-mechanical model.

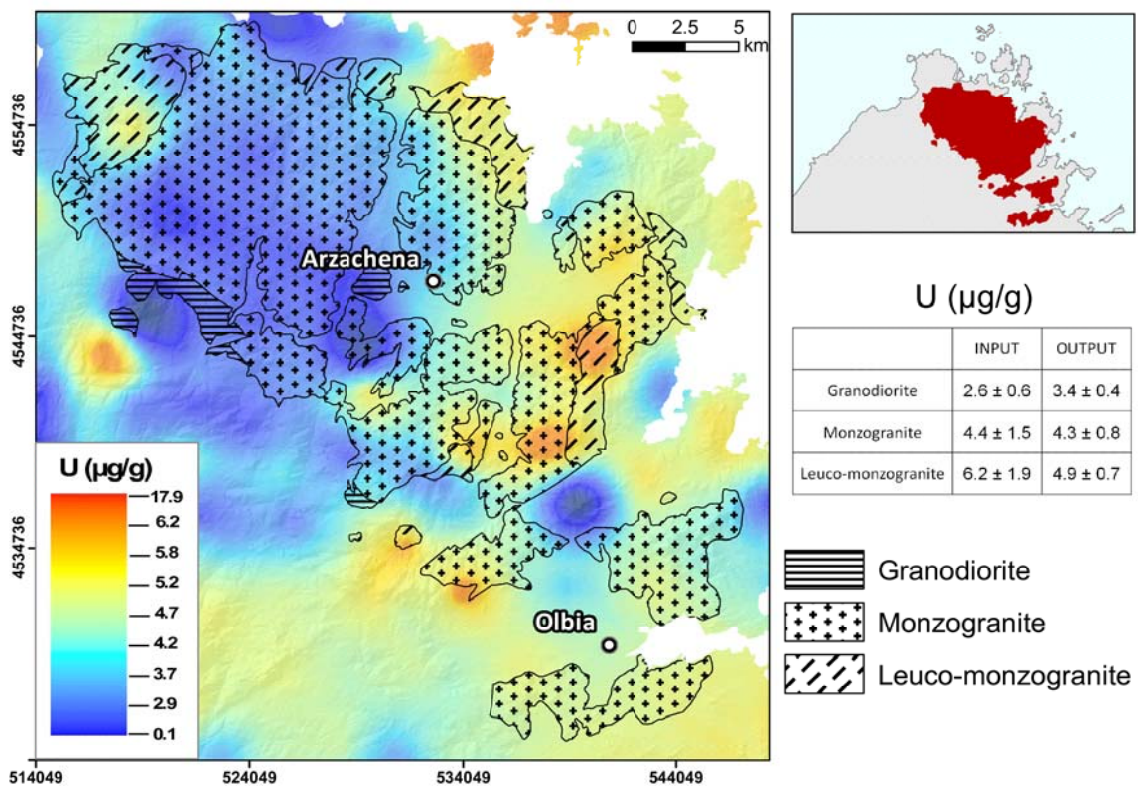

**Figure 6** - The eU distribution map in the Arzachena pluton with the cartographic boundaries of the three different magmatic bodies. The mean and 1σ uncertainty of eU abundances of the input data and the output model are reported for each petrological association (cartographic reference system WGS84 UTM ZONE 32N).

# 6. Conclusions

In this study, we present a 1:100,000 scale map of the eU abundance distribution in the Variscan Basement, which occupies a total area of 2100 km$^2$ in Northeastern Sardinia. The spatial model obtained performing Kriging on the 535 gamma-ray measurements is reported together with the uncertainties of the estimations.

Following a detailed statistical analysis performed on the input dataset consisting of 167 laboratory measurements (HPGe) and 368 in situ measurements (NaI(Tl)), the spatial variability of the eU abundance was studied with the computation and modeling of an omnidirectional ESV (9 lags of 2.2 km). The map was obtained using the Kriging with Variance of Measurement Error method, a geostatistical tool that allows for combining eU abundances with different levels of confidence associated with the two different gamma-ray spectroscopy methods.

The map shows the highest values in the granitoids of the La Maddalena pluton, where the eU content ranges between 6.2 and 9.3 µg/g. However, in the metamorphic basement, outcropping in the southwest of the study area (**Figure 1**), the presence of eclogites intruded in the migmatitic orthogneisses contributes to the lowest uranium abundances (< 2 µg/g).

We suggest that the distribution of the eU content is related to the Post-Variscan brittle structure, reported in **(Casini, Cuccuru, Maino, et al., 2015)**, that affects the metamorphic units and the plutons of the VBNS. In particular, in the Tempio Pausania pluton (southwest of the study area), the major NE-SW faults correspond to anomalies in the eU distribution and mark an area with higher eU content (5 - 7 µg/g) compared to the adjacent sectors of the pluton with lower eU content (~ 3.5 µg/g).

In the AZN, the eU content increases in the more differentiated rocks. This behavior can be verified in the spatial model, even if the ranges of the estimated values for granodiorites, monzogranites and leuco-monzogranites are affected by the 'smoothing effect' typically associated with the Kriging method.

The presented map, integrated with available and more detailed geological maps, is a useful tool for studying the assembly of the intruded plutons and the relationships between the different petrological associations, based on the eU behavior during crustal magmatic processes. Indeed, the study of the eU distribution in the VBNS, particularly in the AZN, could help to refine existing models explaining the post-collisional magmatic processes of the southern European Variscides.

We emphasize that about 90% of the territory of the VBNS is characterized by eU concentrations higher than the average upper continental crust abundance (2.7 µg/g) **(Rudnick & Gao, 2003)**. Since Uranium rich rocks are generally the main source of radon, the presented distribution is a primary criterion for identifying radon-prone areas. Although radon migration depends on many geophysical parameters (e.g. porosity, fractures and permeability of rocks), the assessment of radon gas emission from the underlying bedrock is strongly recommended for mapping the Radon risk, which could be relevant in the coastal areas where the tourism enhances the population density especially in the summer.

## *Acknowledgements*

The authors are extremely grateful to L. Carmignani, T. Colonna, P. Conti and W.F. McDonough who provided insight and expertise that greatly assisted the research. The authors would also like to thank B. Ricci and E. Lisi for comments that considerably improved the manuscript.

This work was partially supported by the Italian Istituto Nazionale di Fisica Nucleare (INFN) through the ITALian RADioactivity project (ITALRAD) and the research grant Theoretical Astroparticle Physics number 2012CPPYP7 under the program PRIN 2012 funded by the Ministero dell'Istruzione, Università e della Ricerca (MIUR).

## *Funding*

This work was partially supported by the Italian Istituto Nazionale di Fisica Nucleare (INFN) through the ITALian RADioactivity project (ITALRAD) and the research grant Theoretical Astroparticle Physics number 2012CPPYP7 under the program PRIN 2012 funded by the Ministero dell'Istruzione, Università e della Ricerca (MIUR).

# URANIUM DISTRIBUTION IN THE VARISCAN BASEMENT OF NORTHEASTERN SARDINIA

Kaçeli Xhixha M.[a,b], Albèri M.[c,e], Baldoncini M.[b,c,e], Bezzon G.P.[d], Buso G.P.[d], Callegari I.[b,d], Casini L.[f], Cuccuru S.[f], Fiorentini G.[c,e], Guastaldi E.[b,g], Mantovani F.[c,e], Mou L.[d], Oggiano G.[f], Puccini A.[f], Rossi Alvarez C.[d], Strati V.[b,c,d], Xhixha G.[b,d], Zanon A[d].

[a] University "Aleksandër Moisiu" Durrës, Department of Engineering Sciences, Faculty of Professional Studies, Str. Currila 1, 2000 - Durrës, Albania.
[b] GeoExplorer Impresa Sociale S.r.l., Via E. Vezzosi, 15, 52100 - Arezzo, Italy.
[c] University of Ferrara, Department of Physics and Earth Sciences, Via Saragat 1, 44122 - Ferrara, Italy.
[d] INFN, Legnaro National Laboratories, Viale dell'Università, 2, 35020 - Legnaro, Padova, Italy.
[e] INFN, Ferrara Section, Via Saragat 1, 44122 - Ferrara, Italy.
[f] University of Sassari, Nature and Environment Department, Via Piandanna 4, 07100 - Sassari, Italy.
[g] University of Siena, Center for GeoTechonologies, Via Vetri Vecchi 34, 52027 - San Giovanni Valdarno, Arezzo, Italy.

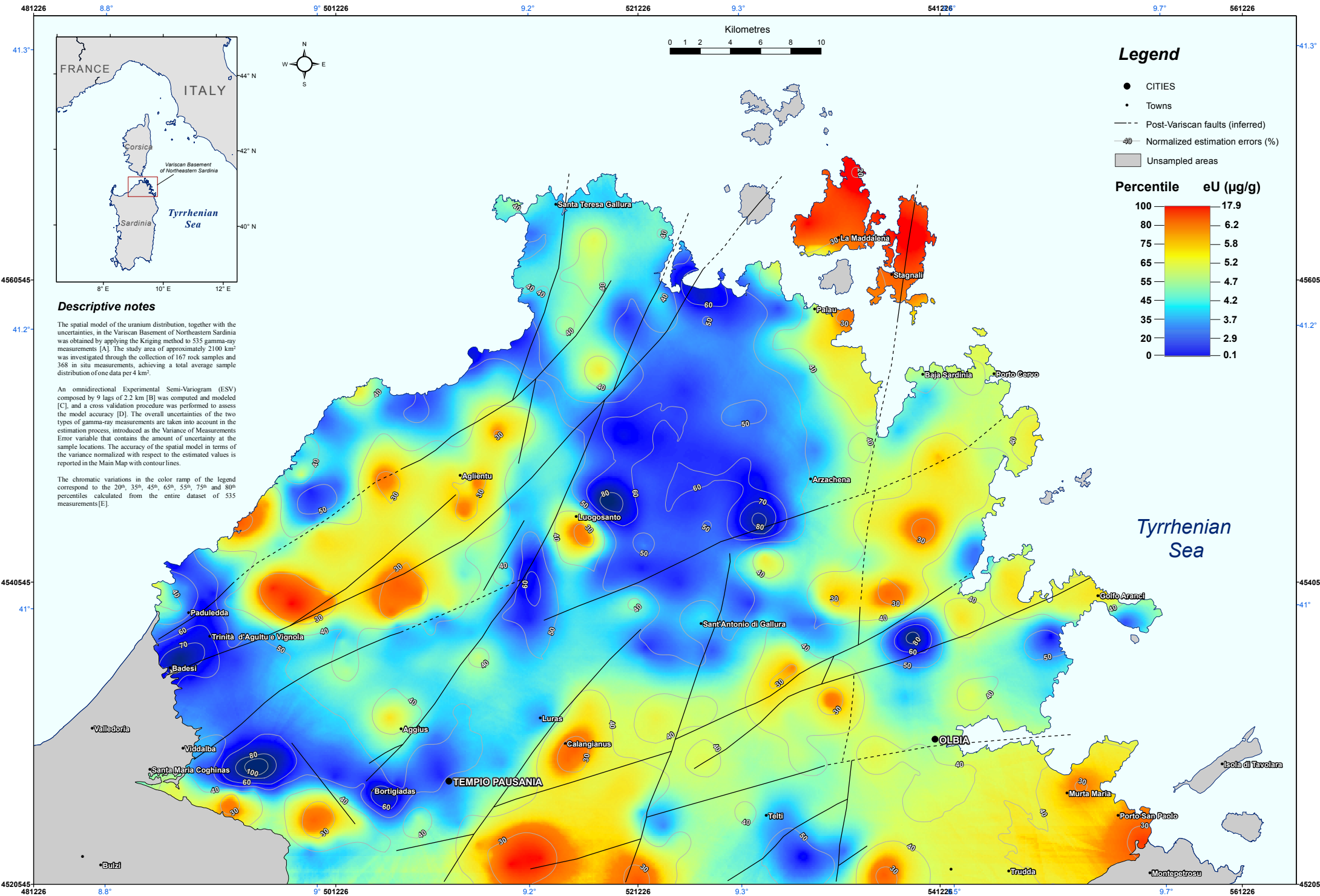

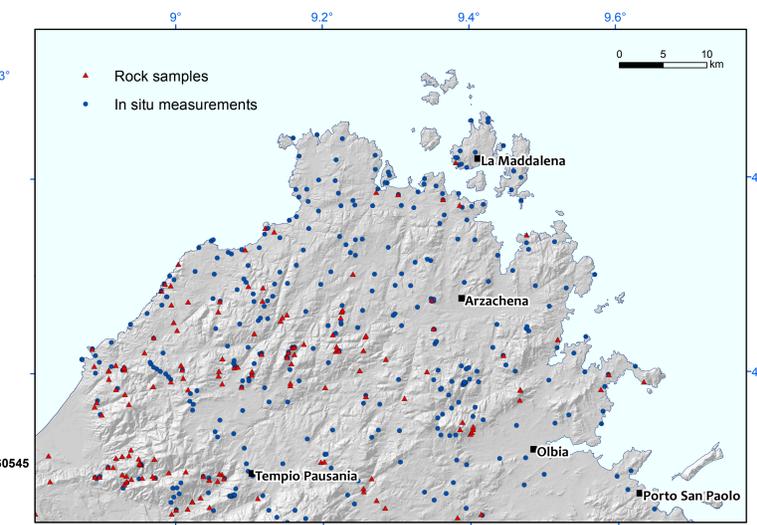

### Descriptive notes

The spatial model of the uranium distribution, together with the uncertainties, in the Variscan Basement of Northeastern Sardinia was obtained by applying the Kriging method to 535 gamma-ray measurements [A]. The study area of approximately 2100 km$^2$ was investigated through the collection of 167 rock samples and 368 in situ measurements, achieving a total average sample distribution of one data per 4 km$^2$.

An omnidirectional Experimental Semi-Variogram (ESV) composed by 9 lags of 2.2 km [B] was computed and modeled [C], and a cross validation procedure was performed to assess the model accuracy [D]. The overall uncertainties of the two types of gamma-ray measurements are taken into account in the estimation process, introduced as the Variance of Measurements Error variable that contains the amount of uncertainty at the sample locations. The accuracy of the spatial model in terms of the variance normalized with respect to the estimated values is reported in the Main Map with contour lines.

The chromatic variations in the color ramp of the legend correspond to the 20th, 35th, 45th, 65th, 55th, 75th and 80th percentiles calculated from the entire dataset of 535 measurements [E].

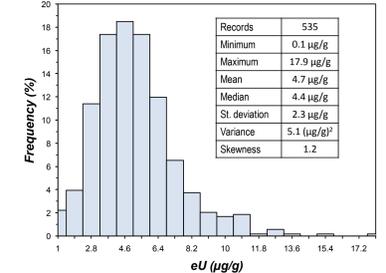

A) Frequency distribution input

| Records | 535 |
| --- | --- |
| Minimum | 0.1 µg/g |
| Maximum | 17.9 µg/g |
| Mean | 4.7 µg/g |
| Median | 4.4 µg/g |
| St. deviation | 2.3 µg/g |
| Variance | 5.1 (µg/g)$^2$ |
| Skewness | 1.2 |

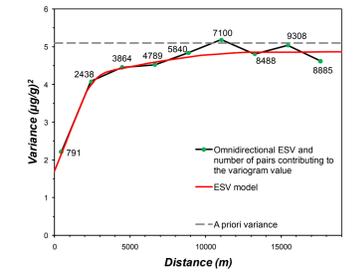

B) Modeling the ESV

C) Parameters of the model

| Structures of variability | Range [km] | Sill [µg/g]$^2$ |
| --- | --- | --- |
| Nugget effect model | - | 1.7 |
| 1st Spherical model | 3.4 | 2.4 |
| 2nd Spherical model | 1.4 | 0.8 |

D) Cross Validation results

| Mean Standardized Errors | -0.04 |
| --- | --- |
| Variance Standardized Errors | 0.74 |

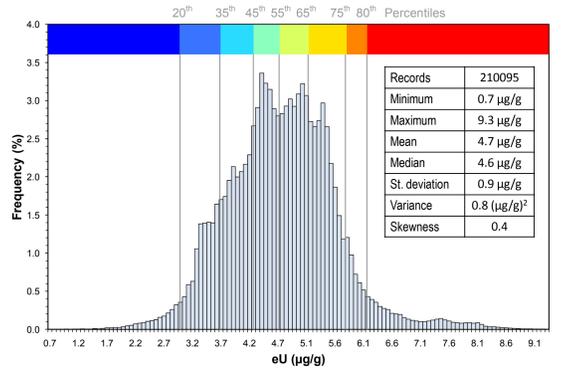

E) Frequency distribution output

| Records | 210095 |
| --- | --- |
| Minimum | 0.7 µg/g |
| Maximum | 9.3 µg/g |
| Mean | 4.7 µg/g |
| Median | 4.6 µg/g |
| St. deviation | 0.9 µg/g |
| Variance | 0.8 (µg/g)$^2$ |
| Skewness | 0.4 |

Scale 1:100,000

CARTOGRAPHIC REFERENCE SYSTEM WGS 84 UTM ZONE 32 - CARTESIAN LATTICE IN BLACK
GEOGRAPHIC COORDINATES REFER TO THE WGS 84 - GEODETIC SYSTEM IN BLUE

© Journal of Maps, 2015